\documentclass[12pt]{article}
\usepackage[polish,english]{babel}
\usepackage[latin1]{inputenc}
\usepackage{times}
\usepackage[T1]{fontenc}
\usepackage{epsfig}

\begin{document}

\title{Spinning Q-balls in the complex signum-Gordon model }

\author{H. Arod\'z, $\;$ J. Karkowski \\ Institute of Physics,
Jagiellonian University\thanks{ Reymonta 4, 30-059 Cracow, Poland}
 \and Z. \'Swierczy\'nski \\ Institute of Computer Science and Computer Methods,\\ Pedagogical
University, Cracow\thanks{Podchor\c{a}\.{z}ych 2, 30-084 Cracow,
Poland }}

\date{$\;$}

\maketitle

\begin{abstract}
Rotational excitations of compact Q-balls in the complex
signum-Gordon model in 2+1 dimensions are investigated. We find that
almost all such spinning Q-balls have the form of a ring of strictly
finite width. In the limit of large angular momentum $M_z$ their
energy is proportional to $|M_z|^{1/5}$.
\end{abstract}

\vspace*{2cm} \noindent PACS: 11.27.+d, 98.80.Cq, 11.10.Lm \\

\pagebreak

\section{ Introduction}

Q-balls belong to the most popular non topological solitons. Formed
by a self interacting scalar field with a $U(1)$ global symmetry,
they are a part of the still largely unexplored world of
non-perturbative phenomena in field theory. As such, for years they
have attracted well justified attention, see \cite{1} for a review
and references. Recent works have been focused mainly on gravitating
Q-balls (also called boson stars) \cite{2}, interactions and
stability \cite{3}, and rotational excitations (spinning Q-balls)
\cite{4}.

The present paper is a sequel to \cite{5}, where non rotating
Q-balls with astonishingly simple analytic form were found in the
complex signum-Gordon model.  The discussed below spinning Q-balls
are their rotational excitations. Study of such excitations is a
natural and desired step in the search for understanding the
nonlinear dynamics of the scalar field.

Our main findings are as follows. First, we present detailed
analytic description of the spinning Q-balls in the planar
signum-Gordon model. It turns out, rather surprisingly, that all
spinning axially symmetric Q-balls except  the ones with the least
non vanishing angular momentum ($|N|=1$) have the form of a ring of
strictly finite width. Outside the ring strip the scalar field has
its exact vacuum value, and inside it the field is given by a
quadratic combination of cylindrical Bessel functions. The inner and
outer radii of the ring are determined from Eqs. (23) below which
can be solved analytically in the limit of high angular momentum. It
is quite remarkable that the signum-Gordon model allows for such a
detailed analytic insight into the structure of the rotationally
excited Q-balls.

In Section 2 below we present certain preliminary material. Section
3 is devoted to explicit  solutions of the field equation. Basic
physical characteristics of the spinning Q-balls are discussed in
Section 4. Several remarks are  collected in Section 5.

\section{Preliminaries }

The Lagrangian of the complex signum-Gordon model has the form
\begin{equation}
L = \partial_{\mu}\psi^* \partial^{\mu} \psi - \lambda |\psi|,
\end{equation}
where $\psi$ is a complex scalar field\footnote{ $^*$ denotes the
complex conjugation, $|\psi|$ is the modulus of $\psi$} in
$(2+1)$-dimensional Minkowski space-time, and $\lambda >0$ is a
coupling constant. The field $\psi$, the space-time coordinates
$x^{\mu}$ and  $\lambda$ are dimensionless -- in physical
applications they have to be multiplied by certain dimensional
constants. The self interaction term $\lambda |\psi|$ regarded as
the function of $(\mbox{Re} \psi, \; \mbox{Im} \psi)$ has the shape
of inverted cone with the tip at $\psi =0$ (the vacuum field). It is
an example of V-shaped field potential. Models with such potentials
have several interesting features \cite{6}, such as compactness of
Q-balls and of other solitonic objects, or a scale invariance of
on-shell type.

Lagrangian (1) is invariant under the global $U(1)$ transformations
$\psi(x) \rightarrow \exp(i \alpha) \psi(x)$ as well as  under
rotations, translations and Lorentz boosts in the $(x^1, x^2)$
plane.   The conserved $U(1)$ charge $Q$, the angular momentum $M_z$
and the energy  $E$ are  given by the following formulas
\begin{equation}
Q = -\frac{i}{2} \int\! d^2x\: (\psi^*\partial_0\psi -
\partial_0\psi^* \psi),
\end{equation}
\begin{equation}
M_z = -\frac{1}{2} \int \! d^2x\:
(\partial_0\psi^*\partial_{\theta}\psi +
\partial_{\theta}\psi^* \partial_0\psi),
\end{equation}
\begin{equation}
E = \int \! d^2x\: (\partial_0\psi^* \partial_0\psi +
\partial_r\psi^*
\partial_r\psi + r^{-2} \partial_{\theta}\psi^*
\partial_{\theta}\psi + \lambda |\psi|),
\end{equation}
where $\theta$ is the azimuthal angle and $r$ the radius in the
$(x^1, x^2)$ plane.

The spinning Q-balls minimize the energy $E$ under the condition
that $Q$ and $M_z$ have fixed values. Introducing Lagrange
multipliers $\lambda_1, \lambda_2$ and the functional $F = E +
\lambda_1 Q + \lambda_2 M_z$, the conditions necessary for the
minimum have the form
\[ \frac{\delta F}{\delta(\partial_0 \psi)} =0 = \frac{\delta F}{\delta(\partial_0
\psi^*)}, \;\;\; \frac{\delta F}{\delta \psi}  = 0 = \frac{\delta
F}{\delta \psi^*}, \] or explicitly
\begin{equation}
\partial_0 \psi + \frac{i}{2} \lambda_1 \psi - \frac{\lambda_2}{2}
\partial_{\theta}  \psi =0,
\end{equation}
and
\begin{equation}
- \triangle  \psi  +  \frac{\lambda}{2} \frac{\psi}{ |\psi|} -
(\frac{i}{2} \lambda_1  + \frac{\lambda_2}{2}
\partial_{\theta}) \partial_0 \psi =0,
\end{equation}
where $\triangle = \partial_r^2 + r^{-1} \partial_r + r^{-2}
\partial_{\theta}^2$. Note that conditions (5), (6) imply that $\psi$ obeys the
Euler-Lagrange equation corresponding to (1):
\begin{equation}
\partial_0^2 \psi - \triangle \psi +  \frac{\lambda}{2} \frac{\psi}{
|\psi|} =0,
\end{equation}
where by definition $\psi/|\psi|=0$ if $\psi=0$, see \cite{5}.

One can show that general solution of condition (5) has the form
\begin{equation}
\psi = \exp(-i\lambda_1 x_0/2) \; \phi(r, \theta + \lambda_2 x_0/2),
\end{equation}
where $\phi$ is an arbitrary (differentiable) function. Axially
symmetric Q-balls obey the condition
\begin{equation}
\psi(r, \theta+\theta_0, x_0)  = \exp(iN \theta_0) \; \psi(r,
\theta, x_0),
\end{equation}
where $\theta_0 \in [0, 2\pi) $ is a rotation angle, $N$ is an
integer. The phase factor $\exp(i N \theta_0)$ is allowed because of
the global $U(1)$ symmetry -- the effect of rotation can be
compensated by the $U(1)$ transformation.  The symmetry condition
(9) together with formula (8) gives
\begin{equation}
\psi = \exp(- i \omega  x_0) \;\exp(iN \theta) \:\chi(r),
\end{equation}
where $ \omega = (\lambda_1 - N \lambda_2)/2. $ The unknown function
$\chi(r)$ obeys the following equation
\begin{equation}
(\partial_r^2 + \frac{1}{r}\partial_r) \chi - \frac{N^2}{r^2} \chi -
\frac{\lambda}{2} \frac{\chi}{|\chi|} = - \omega^2
\chi,
\end{equation}
obtained from Eq. (7) by inserting formula (10). In the case of
rotating Q-balls  $N \neq 0$  and therefore $\chi(0) =0$, otherwise
the function $\psi$ would have a discontinuity at $r=0$.

So far the function $\chi(r)$ can have complex values. In the
intervals of the $r$ variable in which $\chi \neq 0$ we can uniquely
split $\chi(r)$ into the phase and the modulus, $ \chi(r) = \exp(i
G(r)) \; F(r)$, which obey the following equations (obtained from
(11)):
\begin{equation}
\partial_r(r F^2 \partial_r G) =0,
\end{equation}
\begin{equation}
\partial_r^2 F - F (\partial_rG)^2 + \frac{1}{r} \partial_rF -
\frac{N^2}{r^2} F - \frac{\lambda}{2} \mbox{sign}F = - \omega^2 F.
\end{equation}
The $\mbox{sign}$ function has the values $\pm 1$ when $ F \neq 0$
and $ \mbox{sign}(0)  = 0.$ Equation (12) means that $r F^2
\partial_r G $ is constant; substituting $r=0$ gives that constant
equal to 0. Therefore, $G(r)$ has a constant value ($\partial_r
G=0$) in intervals in which $r F(r) \neq0$. For simplicity, we
assume that  $G(r)$ is constant in the whole range   $[0, \infty)$
of the radial coordinate, hence it can be removed by the $U(1)$
transformation. Thus, in the case of simplest axially symmetric
Q-ball with minimal energy
\begin{equation} \psi = \exp(- i \omega x_0) \exp(i N \theta)
F(r),\end{equation} where the nonnegative real function $F(r)$ obeys
the following equation
\begin{equation}
\partial_r^2 F + \frac{1}{r} \partial_r F - \frac{N^2}{r^2} F -
\frac{\lambda}{2} \mbox{sign} F = - \omega^2 F
\end{equation}
with the condition $F(0)=0$.

\section{Explicit form of the profile function $F(r)$}

It is convenient to introduce a new variable $\rho$ and a new
function $f(\rho)$,
\begin{equation}
\rho = |\omega|r, \;\;\; f(\rho) = \frac{2 \omega^2}{\lambda} F(r).
\end{equation}
Then Eq. (15) acquires the parameter free form
\begin{equation}
\partial_{\rho}^2 f + \frac{1}{\rho} \partial_{\rho} f + \left(1- \frac{N^2}{\rho^2}\right) f = \mbox{sign}
f.
\end{equation}

Let us first try the standard tool: a series expansion in a vicinity
of $\rho=0$, $ f(\rho) = \rho^k (a_0 + a_1 \rho + ... )$, where $a_0
\neq 0$.  Because Eq. (17) is invariant under the reflection $f
\rightarrow -f$ we may assume that $a_0 >0$. Then $\mbox{sign} f =
+1$  if we take not too large $\rho$. Eq. (17) considered in the
leading order in $\rho$, represented by terms $\sim \rho^{k-2} $,
implies that $k = |N|$. The next to leading term is then
proportional to $\rho^{|N|-1}$. Therefore, when $|N| > 2$ we do not
find a term $\sim \rho^0$ needed in order to cancel the term
$\mbox{sign} f = +1$. For $|N|=2$ it is the leading term which is
$\sim \rho^0$, but in this case the series expansion also does not
work: Eq. (17) in the order $\sim \rho^0$ gives $0=1$. We conclude
that  the assumed series form of the solution is applicable only
when $|N|=1$. In this case we find that $a_1=1/3$ while $a_0$ is a
free parameter. Strangely enough, the case $|N|=1$ is essentially
different than $|N| >1$.

Luckily, Eq. (17) with $\mbox{sign} f = +1$ coincides with
inhomogeneous Bessel equation for which one can construct the
general solution using a standard method \cite{7}.  The method
requires  Wronskian $W$ of Bessel functions \cite{8},
\[ W = J_{|N|}(\rho) Y_{|N|}'(\rho) - J_{|N|}'(\rho) Y_{|N|}(\rho) =
2/\pi\rho.\] We denote the general solution by $f_+$ in order to
emphasize the fact that it obeys Eq. (17) only in the intervals of
$\rho$ in which $f_+>0$. It has the following form
\begin{eqnarray}
\lefteqn{f_+(\rho) = A J_{|N|}(\rho) + B Y_{|N|}(\rho)}& \nonumber \\
&+ \frac{\pi}{2} Y_{|N|}(\rho) \; \int_{\rho_0}^{\rho} \!d\rho'\:
\rho' J_{|N|}(\rho') - \frac{\pi}{2} J_{|N|}(\rho) \;
\int_{\rho_0}^{\rho} \!d\rho'\: \rho' Y_{|N|}(\rho').
\end{eqnarray}
Here $A,B, \rho_0$ are  constants which are suitably adjusted in
order to satisfy boundary or matching conditions. Further
calculations are carried out separately for $|N|=1$ and $|N| >1$.

\subsection{The $|N|=1$ case}

The condition $ f_+(0)=0$ gives $B=0$ and $\rho_0=0$. Then, formula
(18) gives $a_1 = 1/3$ as expected. It remains to find the constant
$A$ and the interval of $\rho$ in which $ f_+ >0$. Motivated by the
results of \cite{5} we look for compact spinning Q-balls, for which
there exists $\rho_1 >0$ such that $f(\rho)\equiv 0$ if $\rho \geq
\rho_1$ (note that $f\equiv 0$ is a solution of Eq. (17)). The
matching conditions at $\rho=\rho_1$ have the form $\; f_+(\rho_1)
=0, \;\; f_+'(\rho_1) =0\;$. The first condition is just the
continuity of $f$ at $\rho_1$, the second one is obtained by
integrating both sides of Eq. (17) in an infinitesimally  small
interval containing $\rho_1$. The matching conditions  give
\[
A = \frac{\pi}{2} \int_0^{\rho_1}\!\! d\rho \:\rho \: Y_1(\rho),
\;\; \int_0^{\rho_1}\! \! d\rho \:\rho \: J_1(\rho)=0.
\]
The latter condition determines $\rho_1$. Numerically, $\rho_1 =
5.8843...$ and $A = 2.6907...$.  To summarize, the full profile
function in the case $|N|=1$  has the form
\begin{equation}
f(\rho)=\left\{
\begin{array}{lcl} f_1(\rho) & \;\;\; \mbox{if} \;\;\; & 0 \leq \rho \leq \rho_1, \\ 0 & \;\;\; \mbox{if} \;\;\;
&  \rho \geq \rho_1,
\end{array} \right.
\end{equation}
where \[ f_1(\rho) =  \frac{\pi}{2} Y_1(\rho) \; \int_{0}^{\rho}
\!\!d\rho'\: \rho' J_1(\rho') + \frac{\pi}{2} J_1(\rho) \;
\int_{\rho}^{\rho_1}\! \!d\rho'\: \rho' \:Y_1(\rho').
\]

\subsection{The $|N|>1$ case}

The puzzle with the lack of nontrivial series solution close to
$\rho=0 $ is solved at once when we realize that we may take there
the trivial solution $f\equiv 0$. Then the full solution has the
form
\begin{equation}
f(\rho)=\left\{
\begin{array}{lcl}  0 & \;\;\; \mbox{if} \;\;\;
&  0 \leq \rho \leq \rho_0, \\  f_N(\rho) & \;\;\; \mbox{if} \;\;\;
& \rho_0 \leq \rho \leq \rho_1,    \\  0 & \;\;\; \mbox{if} \;\;\; &
\rho \geq \rho_1.
\end{array} \right.
\end{equation}
where  $f_N(\rho)$ is obtained from the general solution (18) by
imposing the matching conditions at $\rho_0$ and $\rho_1$:
\[ f_+(\rho_0) =0, \;\; f_+'(\rho_0) =0, \;\; f_+(\rho_1)=0, \;\;
f_+'(\rho_1) =0. \] The two conditions at $\rho_0$ imply that
$A=B=0$. Thus,
\begin{equation}
f_N(\rho) = \frac{\pi}{2} Y_{|N|}(\rho) \; \int_{\rho_0}^{\rho}
\!d\rho'\: \rho' J_{|N|}(\rho') - \frac{\pi}{2} J_{|N|}(\rho) \;
\int_{\rho_0}^{\rho} \!d\rho'\: \rho' Y_{|N|}(\rho').
\end{equation}

The matching conditions at $\rho_1$ give the following equations for
$\rho_0, \rho_1$:
\begin{equation}
\int_{\rho_0}^{\rho_1}\! \! d\rho \; \rho\:J_{|N|}(\rho) =0, \;\;\;
\int_{\rho_0}^{\rho_1}\!\! d\rho \; \rho\:Y_{|N|}(\rho) =0.
\end{equation}
It is not difficult to determine the radii $\rho_0, \rho_1$
numerically, see Table 1. \vspace{0.3cm}
\begin{center}
\begin{tabular}{|c|c|c|c|c|c|c|c|}
\hline
|N|&1 &2 & 5 &10  &20  &40  &80\\
 \hline
$\rho_0$&0 & 0.6654 &6.4593 & 16.413 &36.392  & 76.382 &156.37\\
 \hline
$\rho_1$& 5.8843&7.7942   & 13.699  & 23.664  & 43.646  & 83.637 & 163.63\\
 \hline
$A(|N|)$&22.082 &39.996& 97.238& 193.72&387.06& 773.95 & 1547.8\\
  \hline
\end{tabular}
\end{center}
Table 1.  Sample of numerical results for $\rho_0, \rho_1, A(|N|)$.
$\rho_0, \rho_1$ have been found directly from purely numerical
solutions of Eq. (17). The integral $A(|N|)=
\int^{\rho_1}_{\rho_0}\!d\rho \:\rho f_N(\rho)$ is considered in
Section 4.  \vspace*{0.3cm}

\noindent Our numerical data show that  $\;\rho_0 \approx 2 |N| -
3.6, \;\; \rho_1 \approx 2 |N| + 3.6\;$ for $|N| \geq 40$.  Such
simple formulas suggest that there exists a simple asymptotic
solution of Eqs. (22) at large values of $|N|$. It turns out that
indeed, this is the case. Let us write Eqs. (22) in the form
\begin{equation}
\int_{\rho_0}^{\rho_1}\!\! d\rho \; \rho \left[J_{|N|}(\rho) + i
Y_{|N|}(\rho) \right] =0,
\end{equation}
and replace the Bessel functions by their asymptotic forms
appropriate for our case in which $\rho_0, \rho_1$ linearly increase
with $|N|$:
\begin{equation}
\left(J_{|N|} + i Y_{|N|}\right)\left(\frac{|N|}{ \cos \beta}\right)
\cong \sqrt{\frac{2}{\pi |N| \tan\beta}} \left[e^{i(|N|\tan\beta -
|N| \beta - \pi/4)} + {\cal O}(N^{-1})\right],
\end{equation}
where $|N| \rightarrow \infty$ and $\beta$ is fixed \cite{8}. Next,
we change the integration variable $\rho$ in (23) to $\beta$ by the
substitution $ \rho = |N|/\cos\beta$. Then, $\rho_0 =
|N|/\cos\beta_0, \; \rho_1 = |N|/\cos\beta_1$. The numerical results
suggest that $\beta_0 = \pi/3 - \delta/|N|, \; \beta_1 = \pi/3 +
\delta/|N|$.  The integration over $\beta$ in the interval
$[\beta_0, \beta_1]$ gives the  equation for $\delta$, namely  $
\exp(3 i \delta) = - 1 + {\cal O}(1/|N|)$. Thus, $\delta = \pi/3 +
{\cal O}(1/|N|)$, and \[ \rho_0 = 2 |N| - \frac{2 \pi}{\sqrt{3}} +
{\cal O}(1/|N|), \;\; \rho_1 = 2 |N| + \frac{2 \pi}{\sqrt{3}} +
{\cal O}(1/|N|),
\]
in agreement with the numerical results $\:(2\pi/\sqrt{3} =
3.6275...)$. Formulas (21), (24) give in the large $|N|$ limit
\[ f_N(\rho) \cong \frac{8}{3} \sin^2[\sqrt{3}\:(\rho-\rho_0)/4] +
{\cal O}(|N|^{-1}).
\]

\section{Physical characteristics of the spinning Q-balls}

The basic physical characteristics of the spinning Q-balls are
calculated from formulas (2-4) in which we insert formula (14):
\begin{equation}
E = \frac{\pi \lambda^2}{2 \omega^4} \int^{\rho_1}_{\rho_0} \!\!
d\rho\: \rho\: [ (\partial_{\rho}f)^2 + f^2 + \frac{N^2}{\rho^2} f^2
+ 2 f],
\end{equation}
\begin{equation}
Q = - \frac{\pi \lambda^2}{2\omega^5} \int^{\rho_1}_{\rho_0} \!\!
d\rho\: \rho f^2, \;\; M_z= - N Q
\end{equation}
($\rho_0 =0$ if $|N|=1$). We see that the angular momentum $M_z$ is
quantized. This has been observed earlier in other models too
\cite{4}.

Formula (25) for the energy can be simplified with the help of two
identities valid for our nonnegative solutions $f(\rho)$ of Eq.
(17):
\[
\int^{\rho_1}_{\rho_0} \!\! d\rho\: \rho \: [ (\partial_{\rho}f)^2 +
\frac{N^2}{\rho^2} f^2 ] =   \int^{\rho_1}_{\rho_0} \!\! d\rho\:
\rho (f^2 -f),   \;\;\; \int^{\rho_1}_{\rho_0} \!\! d\rho\: \rho f^2
=2 \int^{\rho_1}_{\rho_0} \!\! d\rho\: \rho  f.  \] They are
obtained by multiplying Eq. (17) by $\rho f$ or $\rho^2 f'$,
respectively, and integrating by parts. The integral in $E$ is equal
to $5 \int^{\rho_1}_{\rho_0} \! d\rho\: \rho  f_N$, and therefore
\begin{equation}
E= \frac{5\pi \lambda^2}{2 \omega^4} \int^{\rho_1}_{\rho_0} \!\!
d\rho\: \rho f_N = - \frac{5}{2} \omega Q.
\end{equation}
Thus, we need to calculate just one integral $A(|N|) =
\int^{\rho_1}_{\rho_0} \!\! d\rho\: \rho f_N$. It depends only on
$|N|$. Examples of its numerical values are given in Table 1.
Calculation based on the asymptotic formula (24) gives
\begin{equation}
A(|N|) = \frac{32 \pi}{3 \sqrt{3}} |N| ( 1 + {\cal O}(|N|^{-1}).
\end{equation}
It agrees very well with the numerical results already for $|N| \geq
40$.

The two parameters  $\omega$ and $N$ that our solutions contain can
be related to the basic observables, namely
\[ N = - \frac{M_z}{Q}, \;\; \omega = - \frac{2}{5} \frac{E}{Q}.
\]
Furthermore, the first part of formula (27) yields a relation
between the observables, which in the limit of large $|N|$  has the
form
\begin{equation} E \cong 5 \left(\frac{\pi^2}{3\sqrt{3}}\right)^{1/5}
\lambda^{2/5}\: (|M_z| \:|Q|^3)^{1/5}  \end{equation} (note the
absolute values --  $M_z$ and $Q$ can have both signs).

The energy (29) can be regarded as the rest mass of the spinning
Q-ball. Moving Q-balls are obtained by applying Lorentz boosts.

\section{Remarks}

\noindent 1. The field $\psi$ of our  Q-ball solutions reaches
exactly the vacuum value $\psi=0$  at the radius $r_1=
\rho_1/|\omega|$, i.e., at $ r_1 = 2 (3 \sqrt{3}/\pi^2)^{1/5}\:
\lambda^{-2/5}\: |M_z|^{4/5}\: |Q|^{-3/5}$  for large $|M_z|$. If
$|N| \geq 2$ we also have the inner radius $\rho_0 >0$ at which too
$\psi$ reaches the vacuum value exactly. The approach to the vacuum
value is parabolic. For example,  formula (21) gives $f_N(\rho)
\cong (\rho - \rho_1)^2/2 + {\cal O}(\rho - \rho_1)^3$ for $\rho
\rightarrow \rho_1-$. Such behavior of the field is typical for the
models with V-shaped self interactions \cite{6}. It should be
mentioned that similar behavior is observed also in models with a
nonstandard kinetic term (so called K-fields) \cite{9}.

\noindent 2. Above we have discussed  the simplest axially symmetric
spinning Q-balls. One may also consider more general configurations.
First, because the field $\psi$ of the  Q-balls reaches the vacuum
value at the radii $\rho_0, \rho_1$ exactly, one can trivially put
arbitrary number of such Q-balls on the $(x^1, x^2)$ plane, provided
they do not overlap. In particular, one can have Q-balls in the form
of concentric rings, each one with its own values of  $Q$ and $M_z$.
Second, we expect that there exist radially excited versions of the
Q-ball with fixed values of $Q$ and $M_z$. While the profile
function $F(r)$ discussed in Section 3 is non negative, in the case
of the radially excited Q-balls it will have alternating sign before
it parabolically reaches the vacuum value $F=0$. In the case of non
spinning Q-balls this possibility has been shown to exist  \cite{5}.
Third, one may expect that there exist spinning Q-balls which are
not axially symmetric.

\noindent 3. One can easily check that our spinning Q-balls are
stable with respect to radial shrinking or expanding. Nevertheless,
we do not expect that they are absolutely stable because they are
excited states of non spinning Q-balls.  As such, when slightly
perturbed they probably will decay into simpler objects, like
smaller Q-balls (spinning or not), and will emit a packet of
radiation. The dynamics of such processes probably deserves a
separate numerical and analytical investigation.

\section{Acknowledgement}
This work is supported in part by the project SPB nr. 189/6.PRUE/2007/7.

\end{document}